\documentclass[structabstract]{aa}  
\usepackage{graphicx}
\usepackage{txfonts}
\usepackage{color}
\usepackage{enumerate}
\usepackage{lscape}
\usepackage{amssymb}

\begin{document}
   \title{Stellar Cycles from Photometric Data: CoRoT Stars
\thanks{
The CoRoT space mission was developed and  is operated by the French
space agency CNES, with the participation of ESA's RSSD and Science Programmes,
Austria, Belgium, Brazil, Germany, and Spain.}}

\author{ C. E. Ferreira Lopes \inst{1}
\and I. C. Le\~ao \inst{1}
\and D. B. de Freitas \inst{1}
\and B. L. Canto Martins \inst{1}
\and M. Catelan \inst{2,3}
\and J. R. De Medeiros \inst{1}
}

\authorrunning{C. E. Ferreira Lopes et al.}
\titlerunning{Stellar Cycles from Photometric Data: CoRoT Stars}
                 %an outline of variability Amplitude, Periodicity and Rotation}
   \institute{Departamento de F\'isica, Universidade Federal do Rio Grande do Norte, Natal, RN, 59072-970 Brazil\\
              \email{carlos\_eduardo@dfte.ufrn.br}
   \and Instituto de Astrof\'isica, Pontificia Universidad Cat\'olica de Chile, Av. Vicu\~{n}a Mackenna 4860, 782-0436 Macul, Santiago, Chile
   \and Millennium Institute of Astrophysics, Santiago, Chile}
        
   \date{Accepted for publication in A\&A on Aug $25^{st}$ 2015}

% \abstract{}{}{}{}{}
% 5 {} token are mandatory
 
  \abstract
   {Until a few years ago, the amplitude variation in the photometric data had been limitedly explored mainly because of time resolution and photometric sensitivity limitations. This investigation is now possible thanks to the {\em Kepler} and CoRoT databases which provided a unique set of data for studying of the nature of stellar variability cycles. }
   {The present study characterizes the amplitude variation in a sample of main--sequence stars with light curves collected using CoRoT exo--field CCDs.}
   {We analyze potential stellar activity cycles by studying the variability amplitude over small boxes. The cycle periods and amplitudes were computed based on the Lomb-Scargle periodogram, harmonic fits, and visual inspection. As a first application of our approach we have considered the photometric data for 16 CoRoT FGK main sequence stars, revisited during the IRa01, LRa01 and LRa06 CoRoT runs.}
   {The 16 CoRoT stars appear to follow the empirical relations between activity cycle periods ($P_{cyc}$) and the rotation period ($P_{rot}$) found by previous works. In addition to the so-called A (active) and I (inactive) sequences previously identified, there is a possible third sequence, here named S (short-cycles) sequence. However, recovery fractions estimated from simulations suggest that only a half of our sample has confident cycle measurements. Therefore, more study is needed to verify our results and {\em Kepler} data shall be notably useful for such a study. Overall, our procedure provides a key tool for exploring the CoRoT and {\em Kepler} databases to identify and characterize stellar cycle variability. }
   {}


   \maketitle
%
%________________________________________________________________

\section{Introduction}

Dynamo action occurs in the inner stellar layers, but its effects are clearly observed in and above the photosphere, mostly through the magnetic activity cycles. Since the pioneering works by \citet[][]{Wilson-1978} and \citet[][]{Baliunas-1985}, which produced the first large sample of magnetic cycles and rotation information validating stellar dynamo models, an increasing number of studies have been dedicated to determining of empirical relations between the rotation period $P_{rot}$, spectral type, cycle lengths, and other stellar parameters in an attempt to understand the mechanisms of stellar dynamos \cite[e.g.,][]{Noyes-1984,Saar-1992,Baliunas-1995,Brandenburg-1998,Saar-1999,Lorente-2005,Bohm-Vitense-2007}. From these studies we know now that late--type stars exhibit several chromospheric variability behaviors, ranging from regular variations with multi--year periods, similar to the solar cycles, to irregular cycles with no clear pattern and to stars with no long--term variability \citep[][]{Baliunas-1995}. In addition, other studies found shown that short stellar magnetic activity cycles, ranging from 1 to 3 years \cite[][]{Fares-2009,Metcalfe-2010,Sanz-Forcada-2013}. Many authors have focused on the magnetic cycle properties of the Sun; based on sunspot behavior \citep[see][for a comprehensive review]{Hathaway-2010}. Generally, the temporal behavior of sunspots is analyzed using Fourier techniques \citep[e.g.,][]{MacDonald-1989,Kane-1991}, but some investigators have employed analytical methods that are not based on combinations of periodic sine functions \citep[e.g.,][]{Mundt-1991} and some have used wavelet analysis \citep[e.g.,][]{Willson-1999}.

The amplitude variation of stellar photometric data is still poorly studied mainly due to observation length and photometric sensitivity limitations of the observations. Thanks to the {\em Kepler} and CoRoT databases, we now have a unique set of data with which to study the nature of these variations. Recent studies have shown that the photometric variabilities  observed by {\em Kepler} and in data from the Sun behave similarly on all time scales, in the sense that the variability indices ($R_{var}$, defined as the photometric intensity difference between the $5\%$ and $95\%$ values that contain different variability information) are comparable \citep[][]{Basri-2011,Basri-2013}.

\begin{figure*}
\begin{center}
\includegraphics[width=10.cm,height=7cm]{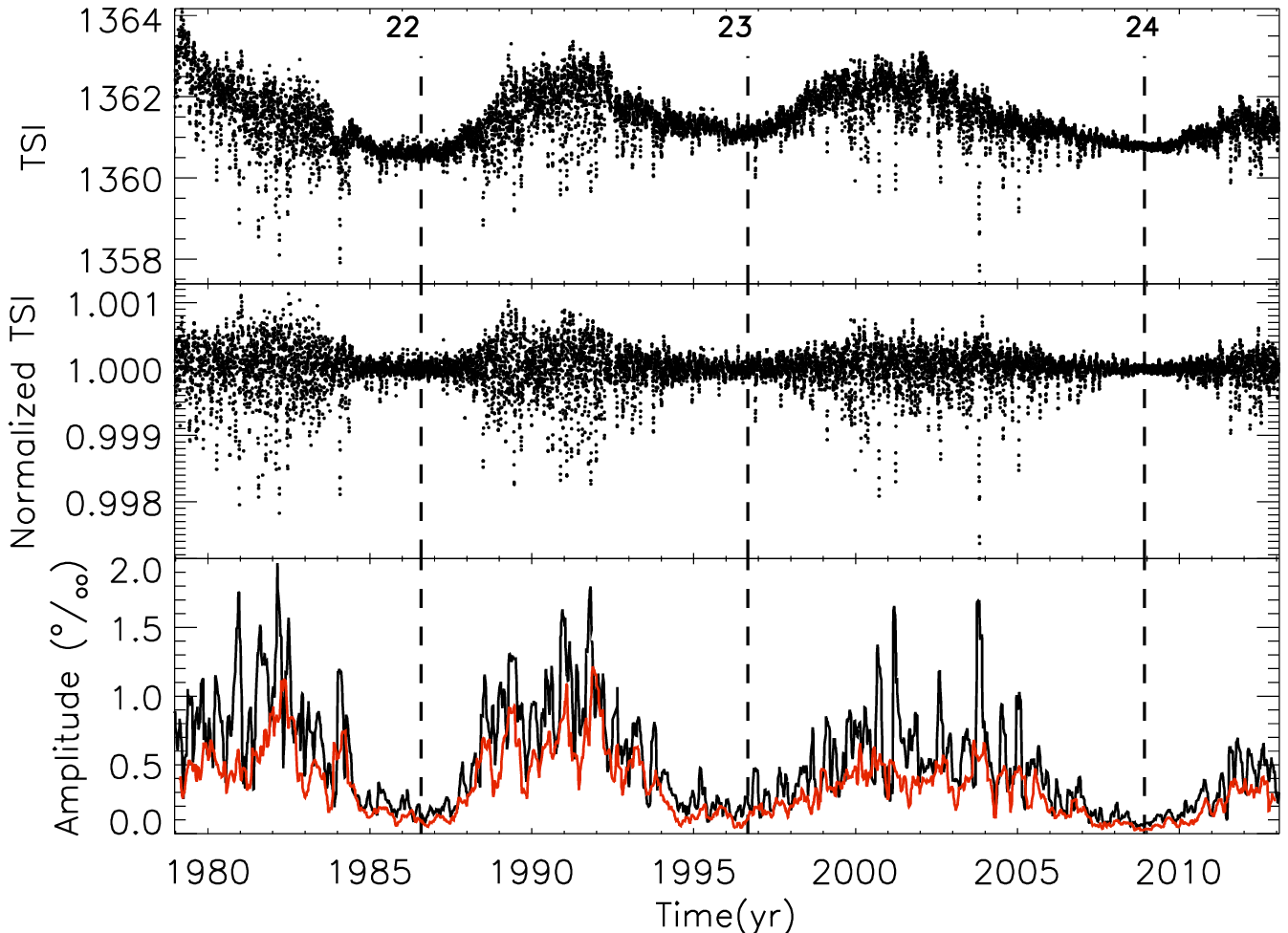}
\includegraphics[width=7cm,height=7cm]{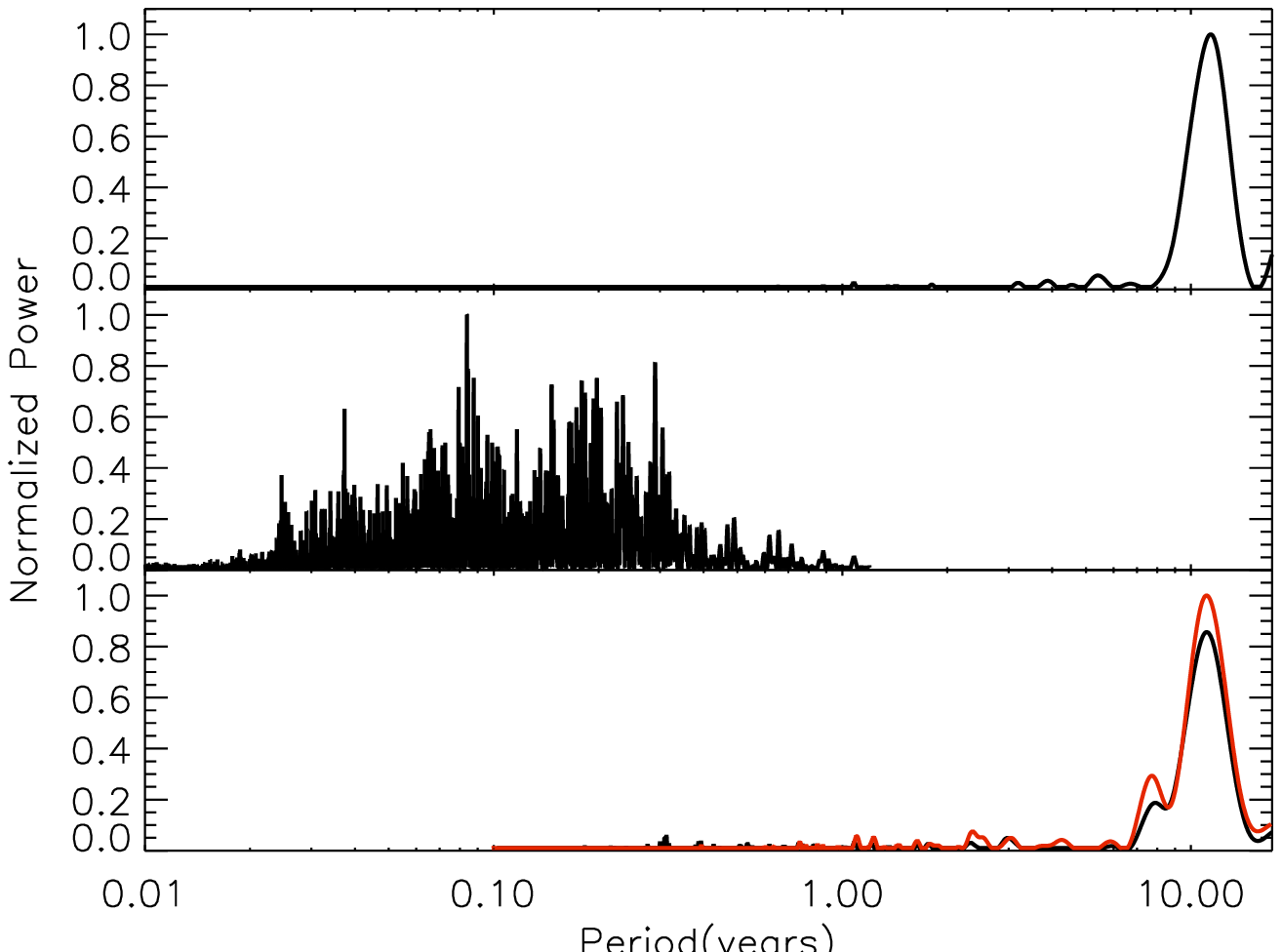}

\caption{The Total Solar Irradiance LC (top-left panel), the prewhitened TSI LC (middle-left panels), the amplitude variation $A(t)$ obtained from the prewhitened TSI LC (bottom-left panel), and Lomb-Scargle periodograms corresponding to each LC (right panels). In the left panels, the dashed vertical lines mark the beginning of each solar cycle, based on review from \cite{Hathaway-2010}. In the bottom-left panel, the $A(t)$ time series is depicted for $N_{box} = 2$ (black solid curve) and $N_{box} = 5$ (red solid curve), whereas in the bottom-right panel the Lomb-Scargle periodogram of $A(t)$ is shown for each $N_{box}$ value with their respective colors.}
\label{brvatsun}
\end{center}
\end{figure*}

More recently, \citet[][]{Mathur-2014} have investigated {\em Kepler} light curves (LCs) of F--type stars searching for indicators of magnetic activity cycles, performing a time--frequency analysis and using the wavelet procedure. In similar fashion, the standard deviation taking in boxes along of LC has been reported as a good activity proxy \citep[][]{Garcia-2014}. As revealed by these  studies, the short and long term behavior of the stellar photometric amplitude variation becomes increasingly important and the actual databases allow us to study this relevant attribute of a variety of star families.

This work aims to present an approach to identify and characterize stellar cycles based on LC variations from data collected by current photometric space missions. The procedure is then applied to a sample of main--sequence CoRoT stars. Section 2 presents a brief description of the method used to determine these cycles, including an application to the Sun's Total Solar Irradiance. The results for CoRoT stars are presented in Sect.~\ref{rescorot}, with conclusions in Sect.~\ref{conclusion}.

%________________________________________________________________

\section{A search for stellar variability cycles}\label{method}
%________________________________________________________________

Solar cycles have been studied from the flux level of solar LCs such as VIRGO data \citep[e.g.,][]{Hathaway-2010}. Activity-related cycles have typically been measured in other stars from the photometric variation of activity indexes related with Ca II emission. Several works used photometric data from the Mt Wilson Ca II H \& K survey of solar-like stars \citep[e.g.,][]{Wilson-1978,Baliunas-1995,Soon-1999,Olah-2000,Messina-2002,Berdyugina-2007,Hall-2007}. Now CoRoT and {\em Kepler} data provide relatively long-term LCs with the best photometric sensitivity and temporal resolution ever. Although long-term variabilities in CoRoT and {\em Kepler} LCs are affected by instrumental long-term trends that hinder physical information, variations such as amplitude are properly retained in the differential LCs.

\citet[][]{Garcia-2010} presented for the first time a method to study long-term stellar cycles using the CoRoT LC of HD 49933 provided in the astero-field channel with a time sampling of 32 s. Based on asteroseismology studies, these authors detected long-term variations of the amplitude, frequency shift, and standard deviation of the modulation produced by the acoustic-mode envelope. These variations were obtained from subseries of 30 days, shifted every 15 days. A similar procedure was recently performed by \citet[][]{Mathur-2013} to study the stars HD 49385, HD 181420, and HD 52265 using CoRoT LCs combined with NARVAL \citep[e.g.,][]{Auriere-2003} spectropolarimetry data. More recently, \citet[][]{Mathur-2014} detected long-term variations in {\em Kepler} LCs of 22 F-stars by considering time-series of different activity proxies, such as the flux standard deviation and temporal sections of the wavelet power spectrum\footnote{This spectrum is somewhat comparable to a time-frequency analysis and consists of a correlation between the time series and a mother wavelet given with different periods along time \citep[][]{Torrence-1998,Mathur-2014}.}. In particular, the standard deviation of LC subseries has been suggested to be a good activity proxy, named $S_{ph,k}$, where $k$ is a multiplier of $P_{rot}$ to define the length of the subseries used to measure the magnetic proxy. As such, \citet[][]{Garcia-2014} also considered this proxy in a study of activity cycles detected in 540 {\em Kepler} pulsating sun-like stars. \citet[][]{Mathur-2014} and \citet[][]{Garcia-2014} have suggested that computing $S_{ph,k}$ within time boxes of $5 \times P_{rot}$, namely $S_{ph,k=5}$, provides a proper detection of Sun-like activity cycles. The activity proxies of those studies are related with the overall LC variability, which for those {\em Kepler} sources are dominated by the rotational modulation.

In the present work, we consider an approach somewhat similar to that performed by \citet[][]{Mathur-2014}. However, we suggest as an appropriate activity proxy the amplitude of the rotational modulation computed within random time boxes, as described below. This proxy was used to compute stellar cycles for a sample of 16 CoRoT stars, with a first application to the Total Solar Irradiance (TSI) LC. The proxy is clearly valid because, for instance, the variability amplitude is somewhat related with the LC standard deviation, as well as with the wavelet power spectrum. In particular, the wavelet analysis has the limitation of requiring a continued time covering to be properly applied. In contrast, our method is an important application for LCs with long-gaps, as detailed in Sect 3.2.

To further elaborate, our procedure is described by the following steps:

\begin{table*}
	\caption{CoRoT stars analyzed in this work.}
  {\centering
  \begin{tabular}{ l c c c c | c c c c c }
  \hline\hline
  \multicolumn{5}{c|}{CoRoT parameters} & \multicolumn{5}{c}{Our parameters} \\
  \hline
CoRoT ID &   ST & LC     &   B      &    V      &  id       &  $P_{rot}$ & $P_{cyc}$ &  Sequence$^a$ & $Q$  \\
                            &         &           & (mag) & (mag)  &            &  (days)      & (days)       &                           &         \\
  \hline
102712791   & G4 & IV & $13.943$  & $13.666$ &  a &    $0.96\pm 0.03$  &  $33\pm 3$   & IS & $\sim 1$ \\
102777006   & G4 & V  & $16.542$  & $15.365$ &  b &    $1.33\pm 0.02$  &  $101\pm 8$  & I & $\sim 1$ \\
102750723   & G1 & V  & $15.162$  & $14.019$ &  c &    $1.44\pm 0.02$  &  $106\pm 7$  & I & $\sim 1$ \\
102721955   & K0 & IV & $14.641$  & $14.210$ &  d &    $2.17\pm 0.06$  &  $187\pm 20$ & I & $0.85$ \\
102752622   & G2 & V  & $14.682$  & $14.208$ &  e &    $2.33\pm 0.07$  &  $346\pm 24$ & AI & $0.59$ \\
102780281   & F8 & V  & $15.889$  & $14.585$ &  f &    $3.0\pm 0.1$    &  $110\pm 8$  & IS & $\sim 1$ \\
102726103   & G4 & IV & $15.096$  & $14.329$ &  g &    $3.7\pm 0.1$    &  $277\pm 21$ & I & $0.64$ \\
102770332   & G7 & V  & $17.644$  & $15.589$ &  h &    $4.2\pm 0.1$    &  $424\pm 41$ & I & $0.42$ \\
102770893   & G5 & V  & $15.169$  & $14.295$ &  i &    $4.3\pm 0.2$    &  $427\pm 45$ & I & $0.42$ \\
102749950   & G1 & V  & $14.883$  & $14.226$ &  j &    $5.4\pm 0.2$    &  $408\pm 26$ & I & $0.42$ \\
102758108   & G5 & V  & $15.098$  & $14.457$ &  k &    $6.1\pm 0.2$    &  $614\pm 55$ & I & $0.34$ \\
102754736   & K2 & V  & $14.825$  & $14.345$ &  l &    $6.9\pm 0.3$    &  $117\pm 8$  & S & $\sim 1$ \\
102723038   & G8 & V  & $15.881$  & $14.477$ &  m &    $8.6\pm 0.5$    &  $210\pm 7$  & S & $0.78$ \\
102720703   & K3 & V  & $14.376$  & $13.862$ &  n &    $10.2\pm 0.6$   &  $650\pm 130$ & I & $0.34$ \\
102778595   & K0 & IV & $15.857$  & $14.700$ &  o &    $11.8\pm 0.7$   &  $201\pm 15$ & S & $0.82$ \\
102738457   & G1 & V  & $15.639$  & $15.047$ &  p &    $12.9\pm 0.6$   &  $239\pm 22$ & S & $0.78$ \\
 \hline\hline
     \end{tabular} \\ }
     \small
     \vspace{0.1in}
     {\bf Note.}\\
     $^a$ Sequence refers to the likely classification of activity level in the $P_{cyc}$ versus $P_{rot}$ diagram (see Sect.~\ref{physics}) based on previous studies \citep[e.g.,][]{Bohm-Vitense-2007,Saar-1999,Brandenburg-1998} (A of active and I of inactive) and in our new results (S of short cycle periods). AI or IS means an intermediary location between A and I or I and S.
  \label{table01}
\end{table*}

\begin{enumerate}

\item For homogeneity, either a TSI or a CoRoT LC is analyzed in normalized flux units, obtained by dividing the original LC by its entire flux average.

\item The LC is then prewhitened. For CoRoT LCs, multiple runs are then combined to provide full time series for each target. This prewhitening must be performed on CoRoT LCs because these data are subject to long--term trends that mask physical signatures at long periods. As such, the TSI data are also prewhitened to simulate the CoRoT LC limitations. Details regarding this prewhitening will be discussed below.

\item We compute $P_{rot}$ as being the main peak of a Lomb-Scargle periodogram \citep[][]{Lomb-1976,Scargle-1982} of the entire prewhitened LC. The periodogram frequency range is defined between $f_0 = 2/T_{\rm tot}$, where $T_{\rm tot}$ is the total baseline of the observations in days, and $f_{N} = 10$ d$^{-1}$, with $10^5$ elements. We refined $P_{rot}$ by analyzing its dispersion defined by Eq. (2) of \citet{Dworetsky-1983} as a function of period. We computed a Gaussian fit with this function in such way that its minimum provided the refined $P_{rot}$ value and its full width at half maximum (FWHM) was assumed to be the $P_{rot}$ uncertainty. More details about this $P_{rot}$ calculation are discussed below. 

\item The LC is analyzed within random time boxes with sizes given by $N_{box}P_{rot}$ with initial times $t_{i}$, where $N_{box}$ is the number of rotation cycles within each box. The choice of the $N_{box}$ value is discussed below.

\item For each box, the phase diagram is computed for a period equal to $P_{rot}$. Then, a harmonic fit is computed as in \citet{De-Medeiros-2013}, given by the following:

      \begin{equation}
          y(t) = \sum_{j = 1}^{M}\left[ a_{j}\sin\left(2\pi f_{1} j t \right) + b_{j}\cos\left(2\pi f_{1} j t \right) \right],
      \label{eq_best_harm}
      \end{equation}

where  $f_{1} = 1/ P_{rot}$, $a_{j}$ and $b_{j}$ are the Fourier coefficients, $M$ is the number of harmonics, and $t$ is the time. For this work, we set $M = 2$, based on a compromise between a reasonable fit and computation time. From this fit, the amplitude $A_i$ is computed as the difference between the maximum and minimum values of $y(t)$.

\item The above steps are repeated in $N$ iterations to obtain a suitable profile of the amplitude variation over time. For this work, we assign $N = 10 T_{tot} / P_{rot}$, which corresponds to approximately 10 amplitude measurements per rotation cycle. This number is chosen in a compromise between the stability of the results and computation time.
 
\item Next, a time series $A(t)$ is defined as being the set of $A_{i}$ values corresponding to each $t_{i}$ value.

\item Finally, the long-term cycle period ($P_{cyc}$) is computed as the main peak of a Lomb-Scargle periodogram of the $A(t)$ time series. For this periodogram, the frequency range is defined between $f_0 = 2/T_{tot}$ and $f_{N} = 1/P_{rot}$, with $10^5$ elements. We also refined $P_{cyc}$ and estimated its uncertainty by following the same procedure as described for $P_{rot}$ in step~3.

\end{enumerate}

\begin{figure*}
  \begin{center}
    \includegraphics[width=8cm,height=5.5cm]{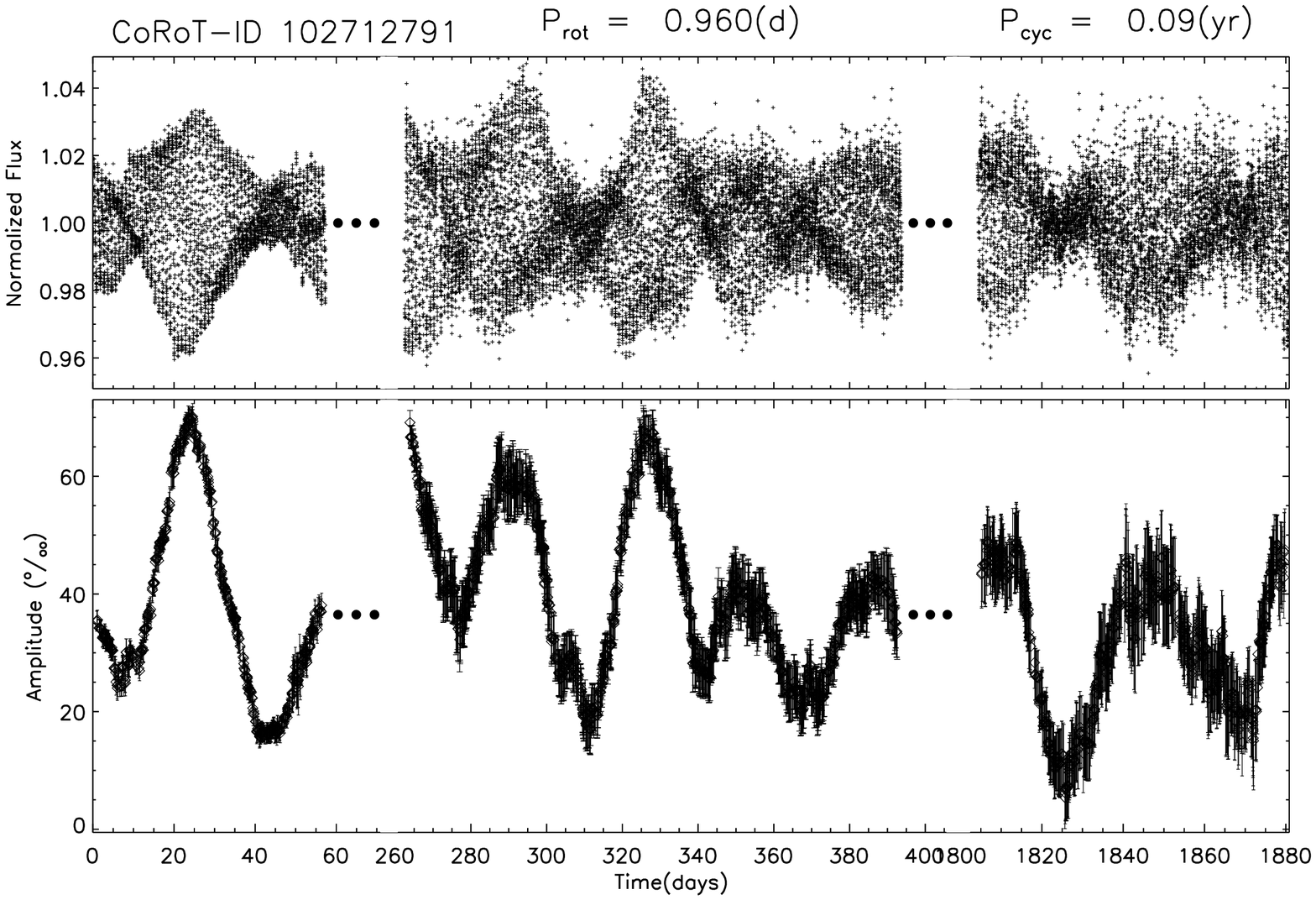}
    \includegraphics[width=8cm,height=5.5cm]{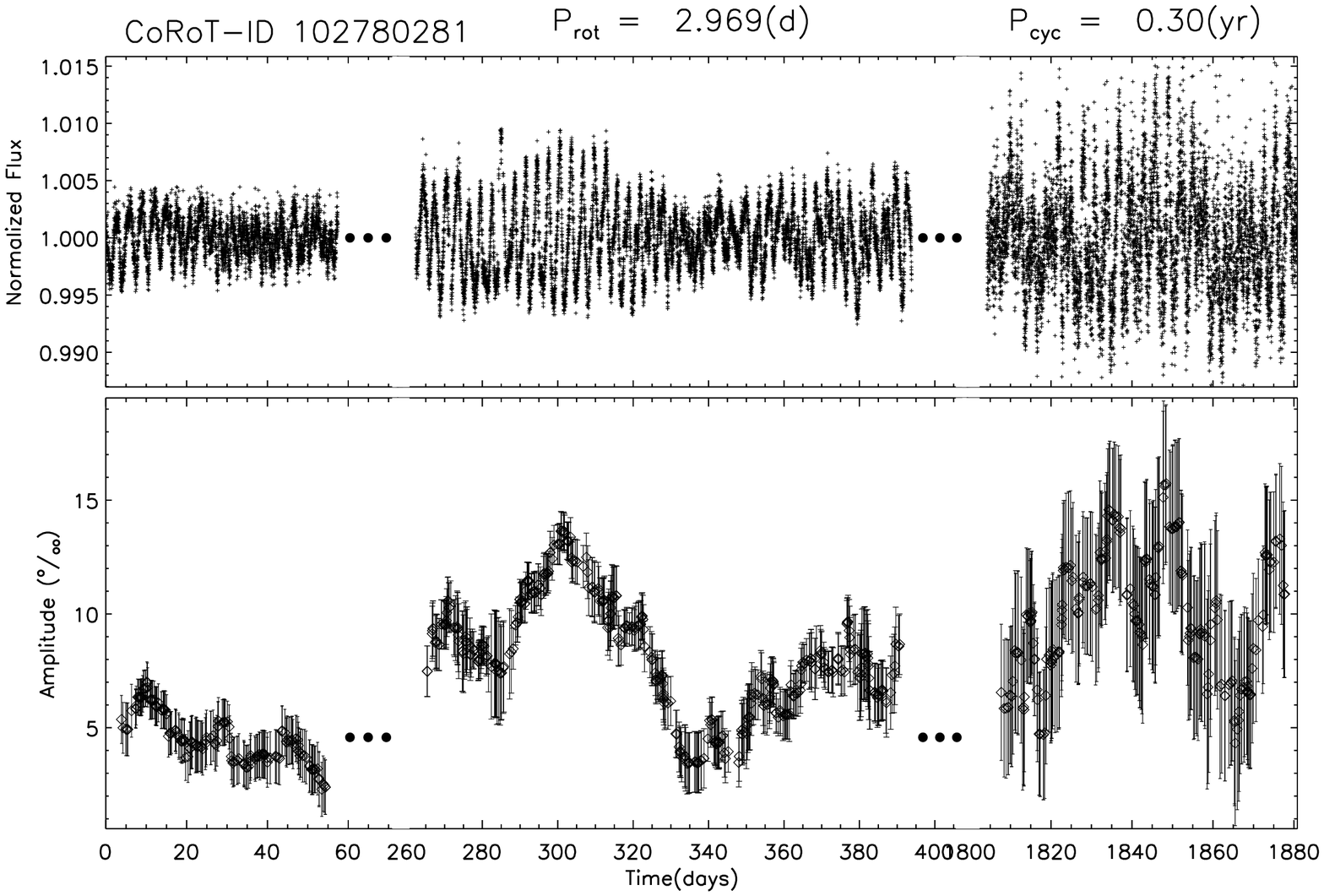}

    \includegraphics[width=8cm,height=5.5cm]{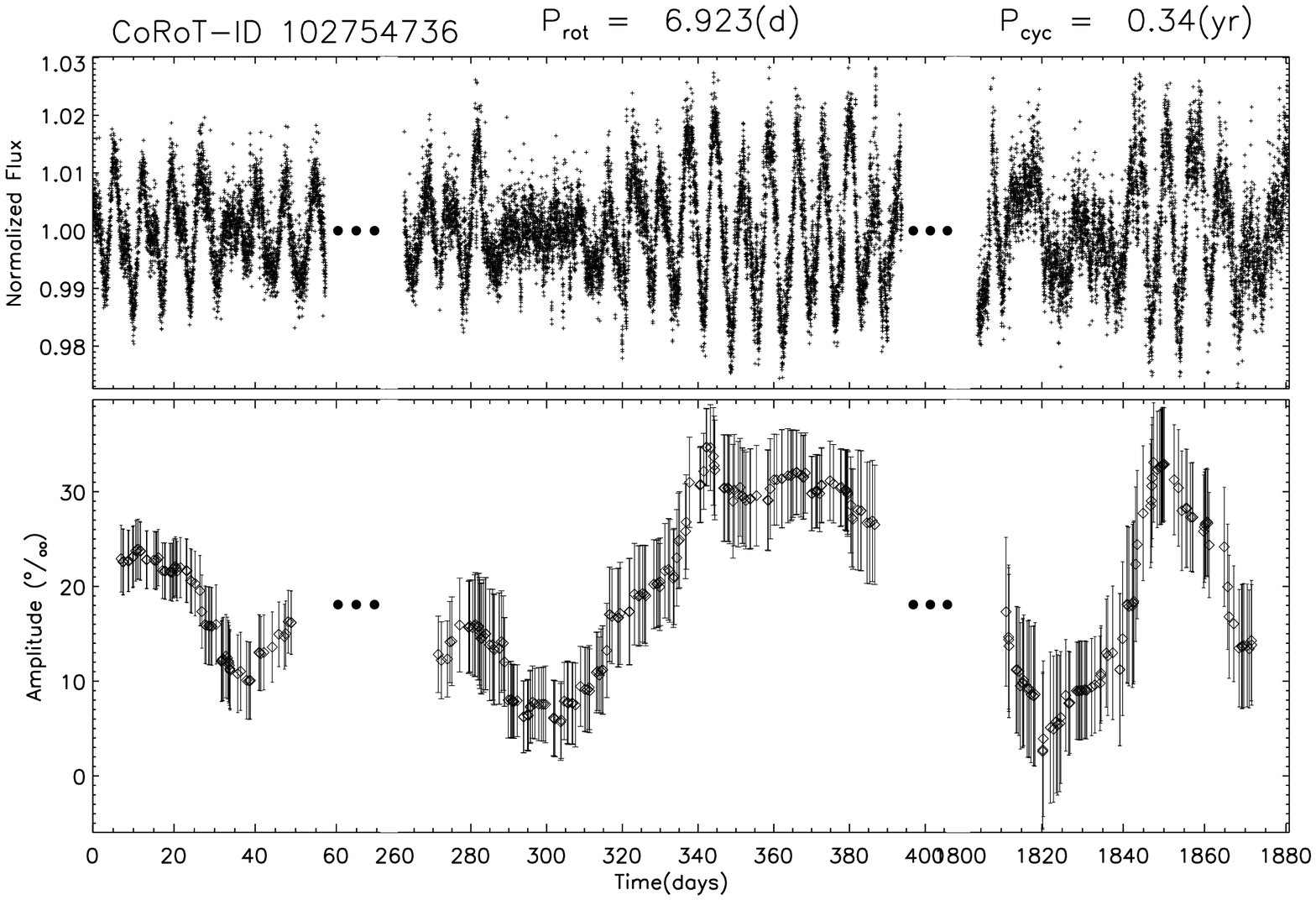}
    \includegraphics[width=8cm,height=5.5cm]{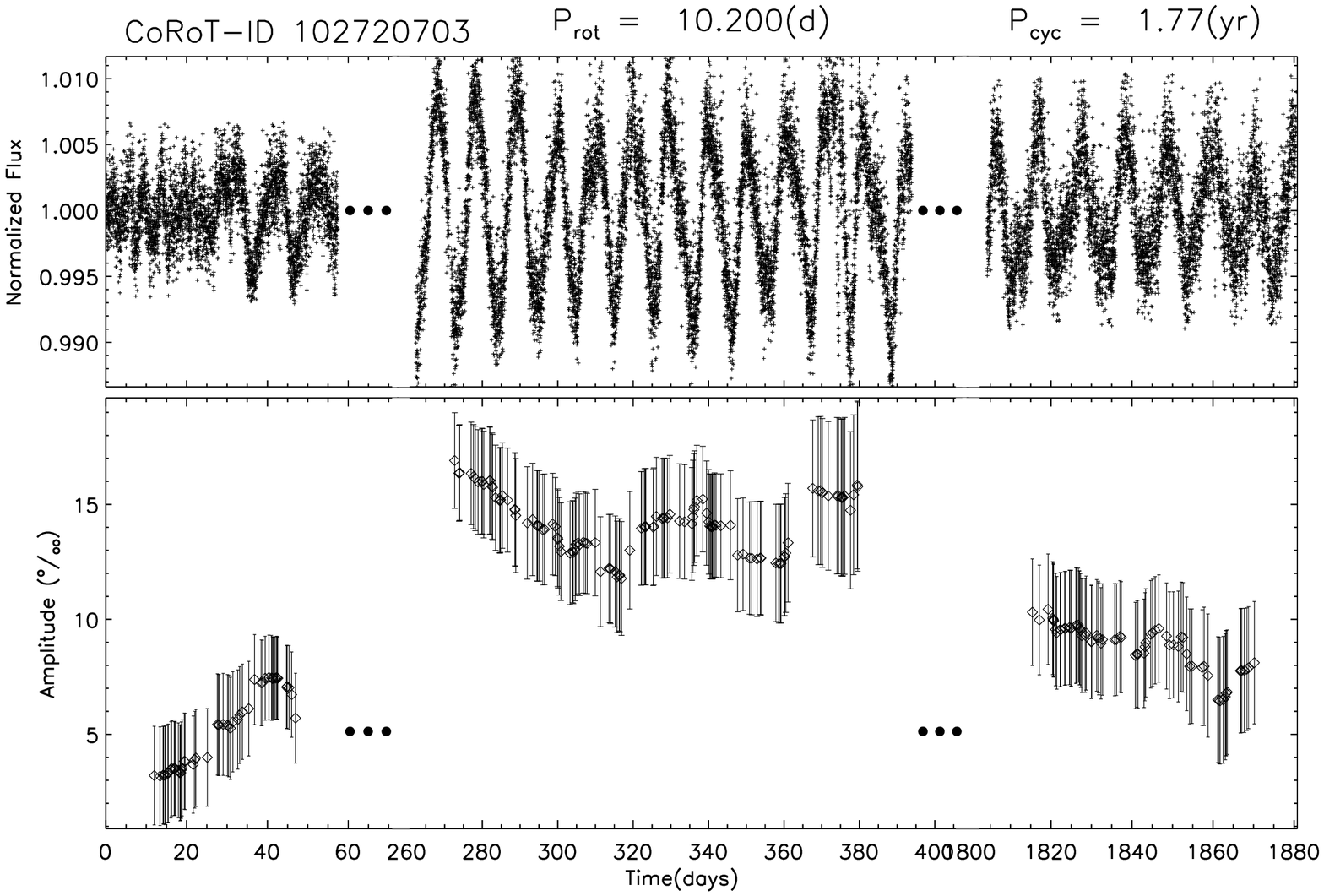}

    \caption{CoRoT LCs and the $A(t)$ time series for typical sources in our final sample. For each star we show the combined LC (upper panel), with the corresponding $A(t)$ time series (bottom panel). The ID and computed $P_{cyc}$ and $P_{rot}$ are given above each panel.}
  \label{lccorot}
  \end{center}
\end{figure*}

Regarding the second step, the prewhitening was tested in various ways and provided stable results. Overall, for CoRoT data, the prewhitening was obtained by dividing the LC of each run by a third-order polynomial fit, as typically performed in the literature \citep[e.g.,][]{Basri-2011,Affer-2012,De-Medeiros-2013}. For TSI data, the prewhitening was obtained by diving each LC by a boxcar smoothed version of itself. The time box in this case can be defined within approximately 100--200 days, corresponding to the typical time windows of the CoRoT runs (of $\sim$150~days). Figure~\ref{brvatsun} (middle-left panel) illustrates a prewhitened TSI LC for a time box of~150~days.

As to the third step, those simple period estimations are valid for approximately $95\%$ of all targets with LCs showing rotational modulation, as estimated in \citet{De-Medeiros-2013}. The remaining $5\%$ typically comprise misdetections of aliases for stars with main spots at opposite longitudes, as explained in the referred work. We also avoided the surface granulation signal, which in the TSI LC is a noisy signature distributed at several frequencies of the power spectrum (see, e.g., Agrain et al. 2004 and references therein). This so-called Solar background is mostly concentrated at low frequencies (typically $\lesssim 0.1$~d$^{-1}$) where the Solar $P_{rot}$ lies. However, that background does not hinder the $P_{rot}$ determination because it has periodogram powers considerably weaker than the peaks produced by the rotating spots. Although we can infer from \citet{Aigrain-2004} that the Solar background may be related with the activity long-term cycles, exploring such a relation is not the scope of our work. As such, for the CoRoT LCs, we assumed the granulation signal to be a noise, which was avoided by selecting LCs with high amplitude-to-noise levels, based on \citet{De-Medeiros-2013}; see also Sect.~3.2.

In the fourth step, the choice for random $t_{i}$ values is not mandatory. However, using such an irregular box sampling minimizes possible window function problems that could be produced by regular bins. In addition, the $N_{box}$ value is not unique. We used $N_{box} = 2$ to have at least two cycles in each box for the calculation of the phase diagram and its amplitude and, at the same time, to have a high sampling in $A(t)$. For validation, we verified that varying $N_{box}$ with slightly greater values provides similar results. Figure~\ref{brvatsun} (bottom panels) illustrates the main data in our analyses for $N_{box} = 2$ (black solid curves) and 5 (red solid curves). Although the amplitude variation plotted in the bottom-left panel is smoother for $N_{box} = 5$ than for $N_{box} = 2$, both values produce similar periodogram peaks, as shown in the bottom-right panels. Therefore, the final results (see Sect. \ref{ressun}) for long time variations are stable over different $N_{box}$ values.

%________________________________________________________________

\section{Results}\label{ressun}
%________________________________________________________________

\subsection{Revisiting the Total Solar Irradiance light curve}\label{tsi}

TSI data have been collected from space over the past 35 years. \citet{Hathaway-2010} presents a chronological list of the instruments that have performed these measurements, with a review of the main results achieved using these data, based on composite TSI time series constructed using different procedures \citep[e.g.,][]{Willson-1997,Willson-1999,Frohlich-1998}. Here, we revisit the TSI LC in normalized flux units to validate the application of our new approach to CoRoT LCs.

Figure~\ref{brvatsun} shows the original and prewhitened TSI LCs (upper and middle left panels, respectively). The Lomb-Scargle periodogram of the original LC (see top-right panel) shows a prominent peak, which is the well-known 11-yr solar cycle. However, as expected, this period disappears in the periodogram of the prewhitened LC (see middle-right panel). Therefore, long period could not be properly detected in other stars from the CoRoT flux variations because of the aforementioned long-term trends. In our new approach, the amplitude variation (bottom-left panel) was obtained from the prewhitened LC and, in this case, the 11-yr solar cycle was confidently recovered in the power spectrum (see bottom right-panel). Therefore, our method is suitable to search for long-term variability cycles with long time span data.

In Fig.~\ref{brvatsun}, the power spectrum of the original TSI LC (see top-right panel) is dominated by long-term variations (including 11-yr solar cycle), whereas the periodogram of the prewhitened LC is dominated by short-term variability (see middle-right panel). As such, we first computed $P_{rot}$ from the prewhitened TSI LC (middle-left panel), as described in Sect.~\ref{method}. The Lomb-Scargle periodogram (middle-right panel) yielded a main period of~$30.6$ days, which correspond to the solar surface $P_{rot}$ at its active latitudes \citep[see, e.g.,][]{Lanza-2003}. From the $A(t)$ time series (lower-left panel), the power spectrum (lower-right panel) recovers well-known long-term periods from the literature: $11.6\pm 0.9$~yr, $8.3\pm0.7$~yr, $1.2\pm0.4$~yr, $2.4\pm0.6$~yr.

The periods of $1.2$~yr and $2.4$~yr computed here are compatible with the typical time scales of $1.1$~yr and $2.8$~yr, respectively, of the so-called Gnevyshev gaps \citep[][]{Gnevyshev-1963,Gnevyshev-1967,Gnevyshev-1977}. These gaps, which are relatively short-term decreases in the solar activity, can be seen in the $A(t)$ time series (Fig.~\ref{brvatsun}, bottom-left panel), especially considering the red solid curve. For instance, during the solar maximum phase of cycle 22, there is a prominent short-term decrease throughout the years 1989--1991 and several fainter decreases are also noticeable during other maximum phases. The $1.2$~yr period computed here is also compatible with a quasi-periodicity of approximately $1.3$~yr, which is related to the stochastic processes of active region emergences produced by the solar magnetic dipole moment and open magnetic flux \citep[][]{Wang-2003}. Another periodicity of approximately $1.0$~yr is also produced by dynamical processes of differential rotation, meridional flow, and supergranule diffusion \citep[][]{Hathaway-2010}. In contrast, \citet[][]{Fletcher-2010} suggests that a period of $2.0$~yr, which is most prominent during the solar maximum phases, may be related to the acoustic modes of the main 11-yr solar cycle. Hence, several phenomena are related to these different activity cycles identified in the TSI LC, such as the main 11-yr cycle, as well as the shorter cycles of $1.2\pm0.4$~yr and $2.4\pm0.6$~yr computed here (1--2-yr cycles henceforth). Therefore, detecting either 11-yr- or 1--2-yr-like cycles in other stars is an important goal for studying activity cycles.

\begin{figure}
  \begin{center}
    \includegraphics[width=4.cm,height=3.5cm]{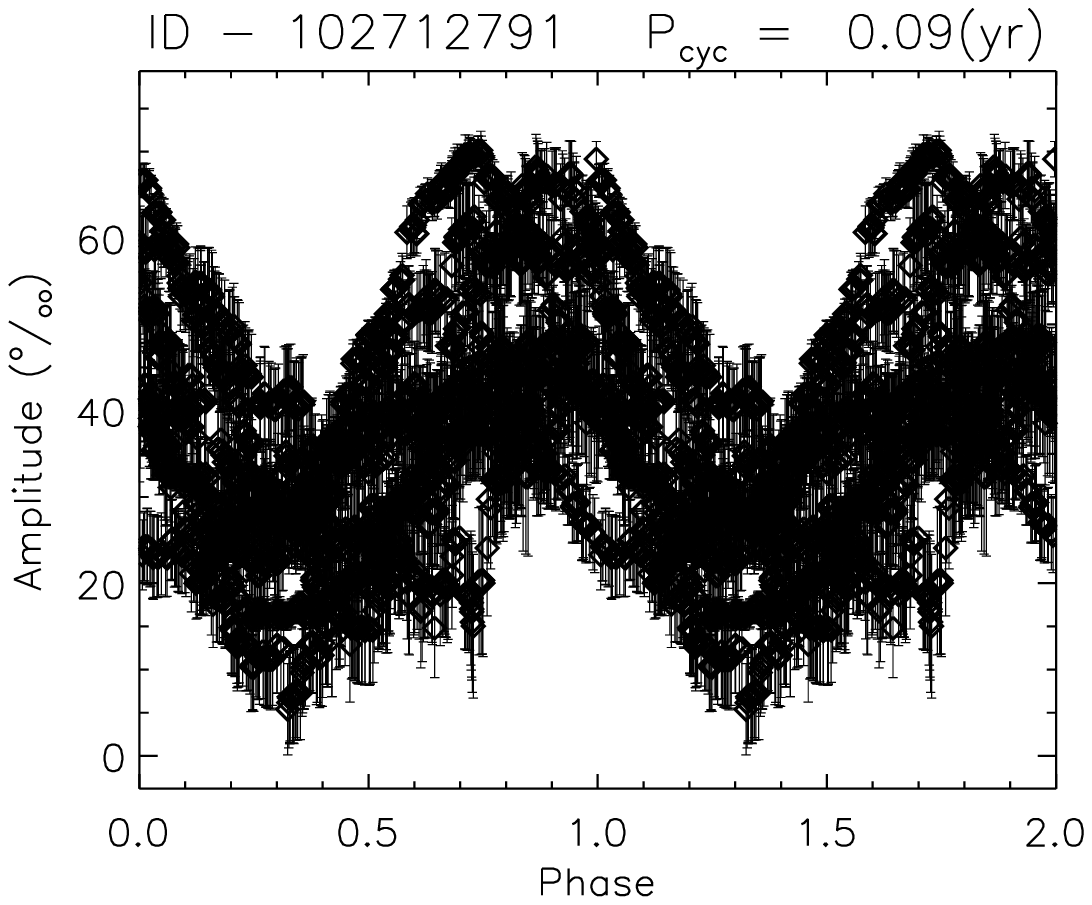}
    \includegraphics[width=4.cm,height=3.5cm]{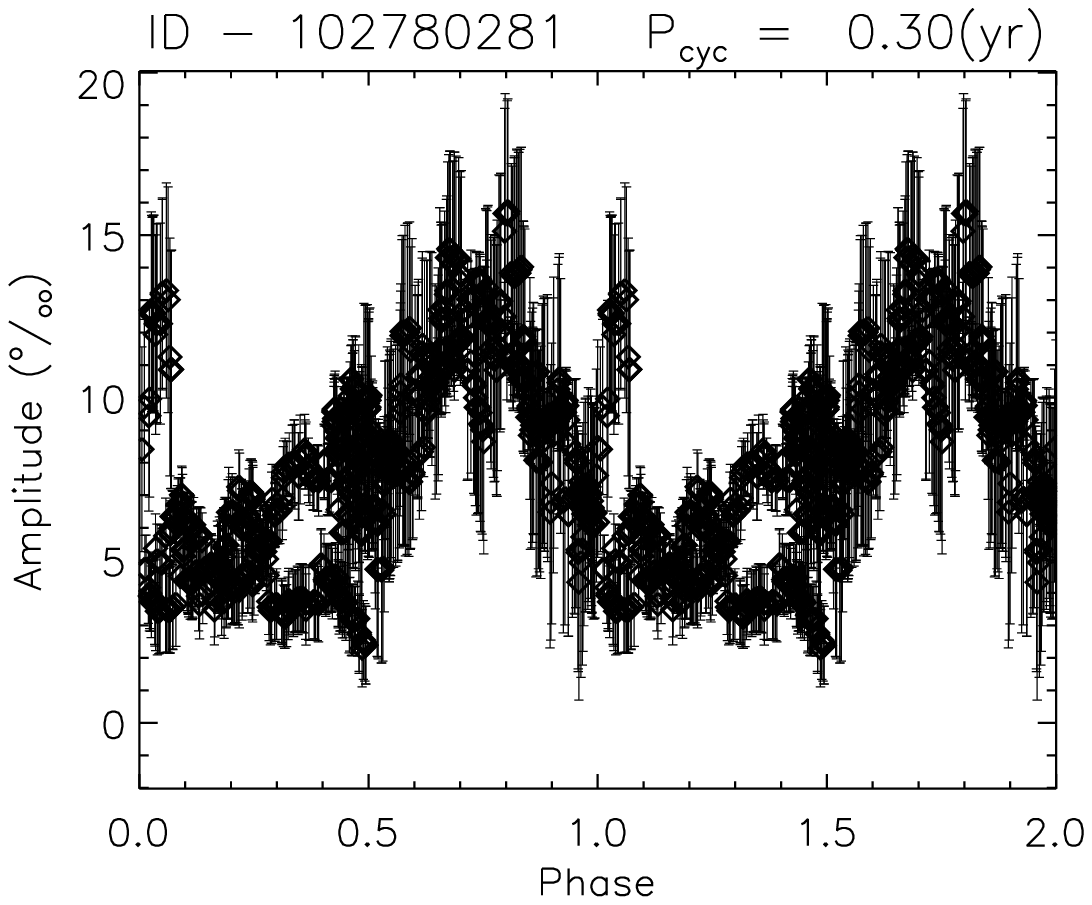}

    \includegraphics[width=4.cm,height=3.5cm]{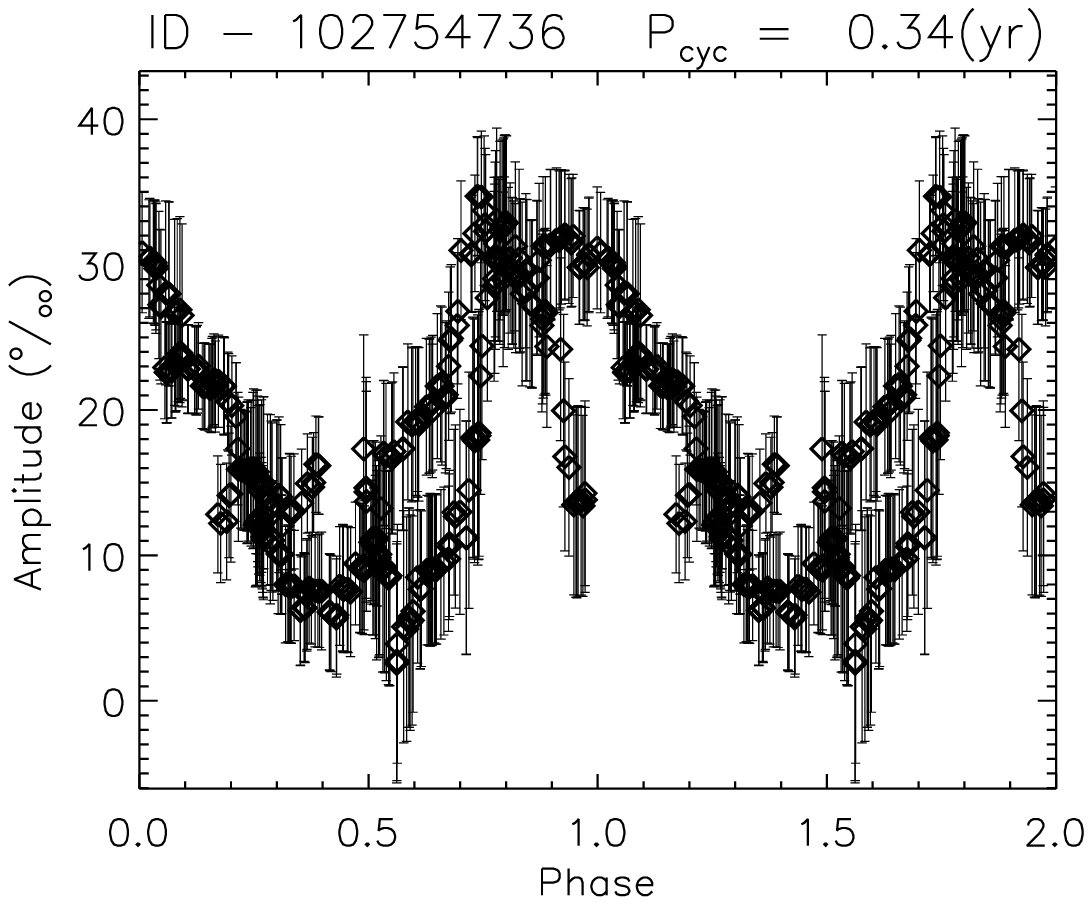}
    \includegraphics[width=4.cm,height=3.5cm]{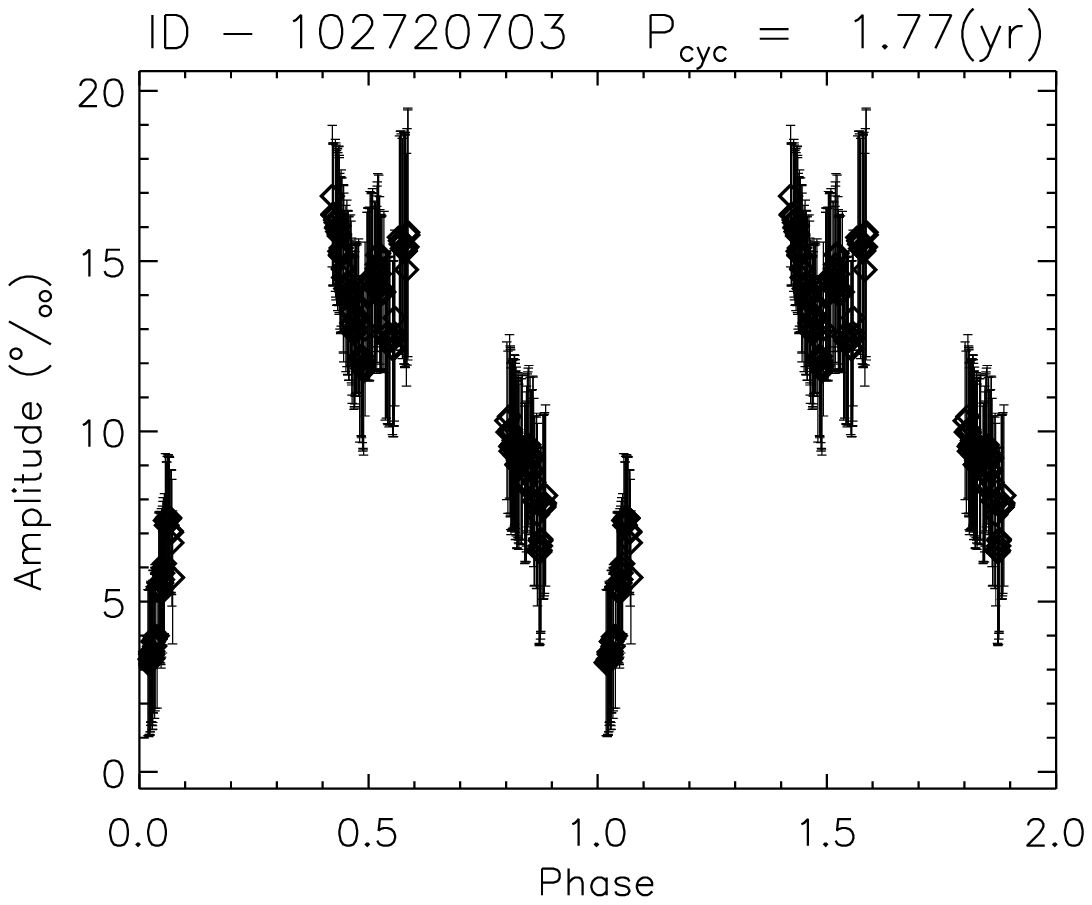}

    \caption{Phase diagrams from $A(t)$ for the stars referred to in Fig.~\ref{lccorot}, with the corresponding ID indicated in the  header.}
  \label{phasediag}
  \end{center}
\end{figure}

%________________________________________________________________
\subsection{Stellar cycles from CoRoT light curves}\label{rescorot}
%________________________________________________________________

To identify stellar cycles in other stars, we analyzed 877 CoRoT LCs from the exoplanet channel, which are publicly available in a reduced format obtained by the so-called CoRoT N1-N2 pipeline\footnote{http://idoc-corot.ias.u-psud.fr/} \citep[see, e.g.,][]{Auvergne-2009}. We selected the LCs that provide the longest combined time coverage, namely those comprising the observing runs IRa01, LRa01, and LRa06, which have typical time spans of 57, 131 and 76 days, respectively. Thus, these LCs combined for each star have coverages of $\sim 264$~days and a total time span of $\sim 1880$ days ($\sim 5.1$ years). From visual inspection we identified 16 FGK type stars with clear rotation signatures and an amplitude-to-noise ratio, as computed in~\citet{De-Medeiros-2013}, greater than 2. These 16 selected stars are listed in Table~1. In addition, a cross-check with the literature showed that 14 stars of this sub--sample were already known as photometric variables \citep[][]{De-Medeiros-2013,Affer-2012}. Figure~\ref{lccorot} illustrates typical LCs of our sub-sample (upper panels) with their respective $A(t)$ time series (bottom panels). The computed $P_{cyc}$ and $P_{rot}$ values of the entire sample of 16 stars are listed in Table~1, with $P_{rot}$ and $P_{cyc}$ ranging from 0.956 to 12.903 days and from 0.09 to 1.77 years, respectively. The table also provides the recovery fractions $Q$ obtained from simulations, which are described in Sect.~\ref{simul}, as well as target identifiers (id) and their evolutionary sequences A, I or S, which are considered in Sect.~\ref{physics}.

Numerically, the $A(t)$ time series were analyzed in this work using a similar technique as that considered in the analysis of radial velocity time series $RV(t)$. Although the physical information of $A(t)$ is obviously different than that provided by $RV(t)$, the numerical techniques involved also consist of identifying a persistent signal in the phase diagram. Therefore, we only need data points in different phases and levels of activity to determine the stellar cycle periods from $A(t)$ (see Sec.~\ref{simul}). Figure~\ref{phasediag} shows the phase diagrams obtained from the four LCs displayed in Fig.~\ref{lccorot}. These phase diagrams exhibit a well-defined signature by considering data points that can be distributed along separated temporal windows (i.e., time series with gaps). All periods were computed with a false alarm probability (FAP), as defined in \citet{Scargle-1982} and in \citet{Horne-1986}, less than $0.01$ (i.e., a significance level $> 99\%$). We verified from simulations (see Sect.~\ref{simul}) that the periods compatible with long-term cycles have a mean $\log(FAP) \sim -10$, whereas incompatible periods have a mean $\log(FAP) \sim -6$. Despite this difference, in any case the FAP values are lower than 0.01. Hence, these low FAP values indicate that the periods detected in the periodogram maxima are statistically acceptable; however, they can have different natures than actual cycles. We present below an approach to verify the confidence of these periods regarding their compatibility with long-term cycles and considering the CoRoT data limitations.

\begin{figure}
   \centering
   \includegraphics[width=0.45\textwidth]{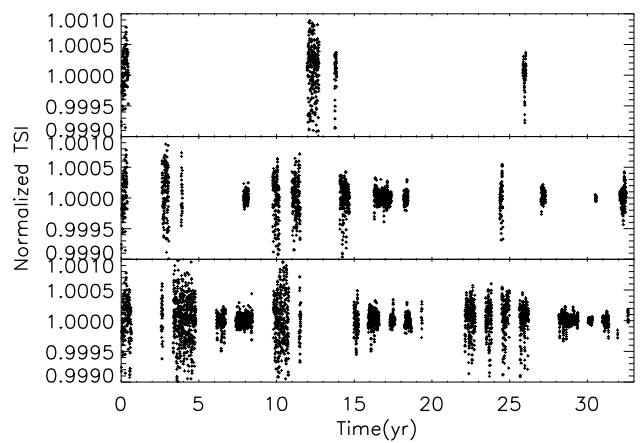} \\
   \caption{Typical examples of the simulated LCs produced from the TSI data. From the top to bottom panels, these LCs have time coverages of approximately 800, 4000, and 8000 days.}
   \label{simlcs}
\end{figure}

\subsection{On the influence of gaps}\label{simul}

The entire LC of each target being combined from multiple runs are not continuous for each target, as described in Sect.~\ref{method}. Thus, the full LCs may exhibit long gaps that may hinder a proper detection of solar-like cycles such as the 11-yr or the 1--2-yr cycles described in Sect.~\ref{tsi}. To test how these gaps may affect the detection of variability cycles in the CoRoT LCs, we performed several simulations by inserting synthetic gaps into the TSI LCs.

\begin{figure}
   \centering
   \includegraphics[width=0.45\textwidth]{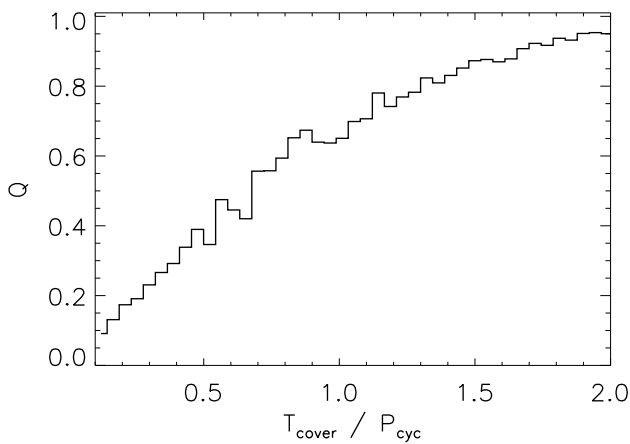} 
   \caption{Recovery fraction $Q$ as a function of $T_{cover}/P_{cyc}$ obtained from $10^4$ simulations where random long-gaps were inserted in the TSI LCs (see text for details).}
   \label{effdiag}
\end{figure}

Specifically, $10^4$ LCs were automatically treated and analyzed as follows. In each simulated LC, which originally cover 35~yr, a certain number of random gaps with durations ranging from 200 to 4000 days was inserted such that time coverages of 400 to 8000 days remained in the simulated LC. Figure~\ref{simlcs} shows typical examples of the resulting simulations with different time coverages. From the entire set of simulations, we computed the recovery fraction $Q$ of the 11-yr cycle as a function of the time coverage to $P_{cyc}$ ratio, or $T_{cover}/P_{cyc}$, as depicted by the black histogram in Fig.~\ref{effdiag}. This calculation was performed within $P_{cyc}$ bins of 0.25~yr by considering that the 11-yr cycle was properly recovered if the main periodogram peak of the simulated LC fell within $11.6 \pm 0.9$~yr, the central value and uncertainty computed from the original TSI LC. A low value of $Q$ does not necessarily means a misdetection, but it means a less reliable measurement of the period. Indeed, the recovery fraction increases if a higher uncertainty is assumed for the 11-yr cycle period, as is the case of some cycles detected in the sample of table~\ref{table01}.

Based on Fig.~\ref{effdiag} and on the relation between $P_{cyc}$ and $P_{rot}$ obtained by \citet{Bohm-Vitense-2007}, we expect to detect 11-yr-like cycles with high confidence ($Q \gtrsim 75\%$) for $P_{rot} \lesssim 2$. This covers the first three or five targets of table~\ref{table01}. Possible 11-yr-like cycles were detected in some targets with $P_{rot} > 2$~d. Even though these cycles have low $Q$ values, we consider them for a discussion in Sect.~\ref{physics} about their overall distribution. The simulations predict high $Q$ values for 1--2-yr-like cycles, suggesting a high confidence for those cycles. However, one should be aware that these cases were assumed as being short versions of the 11-yr cycle within gapped LCs. Estimating $Q$ values for CoRoT data based the own 1--2-yr solar cycles is not obvious because the detection of these cycles in the TSI data is hindered by the strong superimposed 11-yr cycle. In addition, the 1--2-yr cycles have low signals in the TSI LC, in contrast with the possible 1--2-yr-like cycles detected in the CoRoT LCs. Because of these differences between the TSI and CoRoT signals a high $Q$ value can only tell us that a cycle was well-measured because of their long time-coverage, without taking into account other limitations such as the possible hindrance produced by a strong superimposed signal.

\subsection{On the physical nature of the cycle periods}\label{physics}

\begin{figure}
  \begin{center}
  \includegraphics[width=8.cm,height=7.cm]{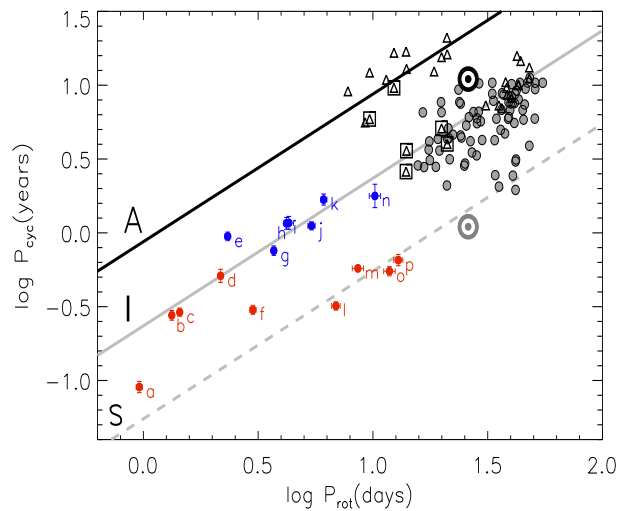}

  \caption{The $P_{cyc}$ (in years) as a function of the $P_{rot}$ (in days) on logarithmic scales, with our data represented by red ($Q > 0.75$) and blue ($Q < 0.75$) circles, data from \citet[][]{Saar-1999} by black triangles and data from  \citet[][]{Lovis-2011} by grey circles. The squares indicate the sources with secondary periods for some stars on the A sequence according to \citet[][]{Saar-1999}. The Sun magnetic cycle periods of $11.6$~yr (black Sun symbol) and $1.1$~yr (gray  Sun symbol) are also represented. The $P_{cyc}$ versus $P_{rot}$ empirical relations (full lines A and I sequences) proposed by \citet[][]{Bohm-Vitense-2007} are also displayed. The empirical relations for shorter term stellar magnetic cycle periods (dashed line S sequence) were obtained by shifting the I sequences by a factor of $1/4$. }
  \label{pcyc_prot}
  \end{center}
\end{figure}

Previous works \citep[e.g.,][]{Bohm-Vitense-2007,Saar-1999,Brandenburg-1998} have identified that $P_{cyc}$ and $P_{rot}$ (as well as the Rossby number) are correlated one another showing an evolutionary behavior with two main sequences: A (active) and I (inactive), which are composed of young and old stars, respectively. Following the approach of \citet{Bohm-Vitense-2007}, we present in Fig.~\ref{pcyc_prot} the relation between $P_{rot}$ and $P_{cyc}$ in logarithmic scale for our CoRoT sample. These are plotted with data from previous works for comparison, as well as with the known parallel A and I sequences. \citet{Bohm-Vitense-2007} also detected a split into the Aa and Ab sequences, which are differed from the I sequence by a factor of $\sim$6 and $\sim$4 (for a given $P_{rot}$), respectively. We consider in Fig.~\ref{pcyc_prot} the sequence Ab. We suggest here to include a third sequence, namely the S sequence, based on the data from CoRoT combined with those from \citet{Lovis-2011}. This S sequence, depicted in Fig.~\ref{pcyc_prot}, parallels the A and I sequences and it differs from the I sequence by a factor of $\sim$1/4 (for a given $P_{rot}$).

The compatibility of the CoRoT data distribution as an extension of the distribution from \citet{Lovis-2011} in Fig.~\ref{pcyc_prot} supports a validation of the S sequence. However, more data is needed for such a final confirmation. Besides, establishing a physical interpretation for this sequence is difficult in the present study because of the limited number of targets and of the CoRoT data limitations. Longer and higher-quality time series provided by the {\em Kepler} mission as already done for a sub-sample of stars by \citet[][]{Mathur-2014}, \citet[][]{Vida-2014} and \citet[][]{Arkhypov-2015} shall help greatly a more in-depth study of the S sequence . Currently we suggest that the S sequence is possibly physical, but it can come from a bias produced by the CoRoT time window limitation, thus exhibiting 1--2-yr-like cycles of stars that would actually lie in the A or I sequence.As such, these short cycles could be physically related with the Solar Gnevyshev gaps. As a second possibility the S sequence could fulfill a third evolutionary region, as \citet{Bohm-Vitense-2007} suggested for the A and I sequences. These possibilities shall be explored in further studies.

There are some stars lying in intermediary regions between two sequences, as depicted in table~\ref{table01}. In particular, targets $a$ and $f$ lie between the I and S sequences and have high $Q$ values. This suggests that their locations in Fig.~\ref{pcyc_prot} are confident and they could be experiencing a transition between two sequences, as we could say about the Sun, which lies between the A and I sequences. A similar case to the Sun could be target $e$. However, we cannot consider its location with confidence because of its relatively low $Q$ value.

%________________________________________________________________
\section{Conclusions and Future Works}\label{conclusion}
%________________________________________________________________

In this study we have shown that the $A(t)$ time series, which expresses the behavior of the amplitude variation as a function of time, seems to enable an important procedure to compute stellar cycles from photometric LCs. In this context, to check our procedure, we have shown that the variability periods for the Sun from the $A(t)$ time series agree with those computed in previous works; that used different approaches. In addition, the shape of the $A(t)$ time series allows us to determine the beginning of the grand minimum and examine the global behavior of the solar magnetic cycle.

From the $A(t)$ time series, we have computed variability cycles and rotation periods for 16 FGK CoRoT stars. The behavior of the obtained cycle periods in the $P_{cyc}$ versus $P_{rot}$ distribution follows a similar relation found in other studies for other families of stars, which suggests a physical nature of the identified stellar cycles. Simulations from Solar TSI data indicate that about a half of our sample has a recovery fraction $Q > 0.75$ and, therefore, further study is needed for more conclusive results. The confirmation of the $P_{cyc}$ versus $P_{rot}$ empirical relation identified in this work combined with those identified in previous studies \citep[e.g.,][]{Bohm-Vitense-2007} may allow us to develop a more robust explanation for their physical nature.

To date, the long-term stellar variability of stellar LCs, such as those collected by the CoRoT and {\em Kepler} space missions, remains largely unexplored. In many cases these stellar LCs may contain large gaps or instrumental trends that hinder identifying long-term physical variations. The current procedure, namely the $A(t)$ time series approach, offers a clear methodology to identify and compute variability cycles from the photometric database of these missions. As such, the longer time span provided by {\em Kepler} and PLATO missions \citep[][]{Rauer-2014}, shall allow us to compute more reliable $P_{cyc}$ measurements and thus a more in-depth study about stellar cycles.

\section*{Acknowledgements}

Research activities of the Observational Stellar Board of the Federal University of Rio Grande do Norte are supported by continuous grants of CNPq and FAPERN Brazilian agencies and by the INCT-INEspa\c{c}o. C. E. F. L. and I. C. L. acknowledge a Post-Doctoral fellowship of the CNPq and CAPES; M.C. acknowledges support from the Chilean Ministry for the Economy, Development, and Tourism's Programa Iniciativa Cient\'{i}fica Milenio through grant P07-021-F, awarded to The Milky Way Millennium Nucleus; from the BASAL Center for Astrophysics and Associated Technologies (PFB-06); from Proyecto Fondecyt Regular \#1110326; and from Proyecto Anillo ACT-86. The authors would like to thank to an anonymous referee for valuable comments and suggestions that improved greatly the present work.

\bibliographystyle{aa}
\bibliography{mylib.bib}

\end{document}